\documentclass[seceq]{ptptex}
\usepackage{wrapft}

\notypesetlogo  

\title{On the Entropy Increase in the Black Hole Formation%

}

\author{
Tetsuya \textsc{Hara}\footnote{E-mail: hara@cc.kyoto-su.ac.jp, 
}, 
 Keita \textsc{Sakai}
,
and Daigo \textsc{Kajiura}\footnote{E-mail: i253037@cc.kyoto-su.ac.jp}%
}

\inst{Department of Physics, Kyoto Sangyo University, Kyoto 603-8555, Japan
\vspace{0.3cm}}

\abst{\hspace{5cm}{\large Abstract}

\vspace{0.2cm}

The entropy increases enormously when a star collapses into a black 
hole.  For the radiation dominated star, the temperature has decreased 
from the almost gravitational equilibrium one to the so-called black 
hole temperature.  If we interpret this temperature decrease as the 
volume increase of the black hole, the entropy increase could be 
understood as the free expansion of radiation through the increased 
volume.  Under this interpretation, it is derived that the entropy is 
proportional to the horizon surface area and the microscopic states of 
the black hole entropy is naturally understood as the statistical 
states of the enlarged phase space. 
}

\begin{document}
\maketitle
 
  Since the research of black hole thermodynamics was born a little over three decades 
ago\cite{rf:BEK, rf:BAR, rf:HAW1}, 
several striking concepts have been accepted. \cite{rf:WAL, rf: FRO, rf: MUK, rf:PAG}
One of them is that the black hole entropy is proportional to the surface area. \cite{rf:KIE,rf:JAC, rf:THO}
However there are still questions about the fundamental concepts of the black hole entropy in 
its microscopic interpretation. \cite{rf:BEK2, rf:HEL, rf:BOU} 
In this letter, we investigate this problem through the point that the entropy has increased enormously 
after the black hole formation.
In the following, units with $c=\hbar=G=k_B=1$ are used, however in some cases the physical constants are expressed.@

  The spherical radiation dominated massive star is considered to collapse due to 
its gravitational instability.  \cite{rf:KIP,rf:CAR, rf:WEI} 
As the gravitational energy is almost equal to the thermal energy, 
 the relation $GM^2/r\simeq \epsilon V_{\star}/3$ is satisfied, where $M$, $V_{\star}$ and $\epsilon$ are the star 
mass, volume and 
radiation energy density, respectively. If we take $M\simeq \epsilon V_{\star}$ ,  the relation
$r \simeq r_{\rm g}=2M$ is derived in the final collapse stage of the black hole formation.  
If it is taken $r\simeq(3M/(4\pi \epsilon))^{1/3} \simeq r_{\rm g}$, 
the temperature $T(=T_{\star})$ is estimated as
$$
T_{\star}\simeq 
{m_{\rm pl} }\left(\frac{m_{\rm pl}}{M}\right)^{1/2} 
,\eqno{(1)}
$$
where $\epsilon =\tilde{a} T_{\star}^4  $ and $m_{\rm pl}=\sqrt{\hbar c/G}$ are taken, 
being $ \tilde{a} =\pi ^2 k_B ^4/(15 \hbar^3 c^3)$.
  It shows  $T_{\star}\propto M^{-1/2} $
and the total entropy $ S \simeq sV_{\star} \simeq 4\tilde{a}T_{\star}^3V_{\star}/3 \simeq 4Mc^2/(3T_{\star})$ is 
expressed by
$$
S\simeq 
\left(\frac{A}{\ell_{\rm pl} ^2}\right)^{3/4}
\simeq10^{57}\left(M/M_{\odot}\right)^{3/2},\eqno{(2)}
$$
where $A=4\pi r_{\rm g} ^2$ and $\ell_{\rm pl}=\sqrt{G\hbar/c^3}$ are the black hole surface area 
and Planck length, respectively.  
It should be noted 
that entropy $S$ is proportional to $A^{3/4}$, before black hole formation.  

On the other hand, the Bekenstein-Hawking temperature and entropy of a black hole are
 $T _{\rm BH}=1/(8 \pi M )$ 
and $S=\intop d(Mc^2)/T_{\rm BH}=M/2T_{\rm BH}\propto A/(4l_{pl}^2) \simeq10^{77}\left(M/M_{\odot}\right)^{2}
$, respectively.  
Consequently the entropy 
has increased by an enormous factor  
$(A/\ell_{\rm pl} ^2)^{1/4}\simeq 10^{20}(M/M_{\odot})^{1/2} $ 
during the black hole formation.

The massive star is to collapse 
even when $r\gg r_{\rm g}$ and, if it is assumed to be adiabatic, the entropy is 
constant during the contraction before black hole formation.  
Because of $S\simeq M/T$ 
before and after the black hole formation, this entropy increase is mainly due to the temperature 
decrease.  Before the black hole formation, the temperature is 
$T_{\star} \propto m_{\rm pl}\thinspace (m_{\rm pl}/M)^{1/2}$ and, after the black hole formation, it becomes 
 $T_{BH} \propto m_{\rm pl}\thinspace(m_{\rm pl}/M)$ .

 If we consider that a black hole is mostly composed of black body radiation, the entropy increase could be understood 
through the relation 
$M\simeq \epsilon V\propto T_{\star}^4 V_{\star}\propto T_{\rm BH}^4 V_{\rm BH}$, 
then the black body temperature is related as 
$T_{\rm BH}\simeq T_{\star}(V_{\star}/V_{\rm BH})^{1/4} \simeq T_{\star}(m_{\rm pl}/M)^{1/2} $, 
where $ V_{\rm BH}$ is the effective 
black hole volume or the effective volume of the gravitationally collapsed object.\cite{rf:SOR}  
For $ T_{\rm BH} \ll T_{\star}$,  $ V_{\rm BH}$ is much greater than $ V_{\star}$ 
as $ V_{\rm BH}\simeq V_{\star}(T_{\star}/T_{\rm BH})^{4} \simeq V_{\star}(M/m_{\rm pl})^{2}(\propto M^5)$.

If the formation process is adiabatic,  the temperature changes as $T \propto (1/V)^{1/3}$ 
 and it is no concern for us, because the entropy does not change.  If the process is free expansion, 
the temperature changes appropriately as 
$T \propto (1/V)^{1/4}$, because the internal energy is constant ($U\propto T^4V =const$), 
and/or, thermodynamically, the following relation is satisfied
$$
\left( \frac{\partial T}{\partial V}\right)_U=- \left( \frac{\partial U}{\partial V}\right)_T {\Big/ }
\left( \frac{\partial U}{\partial T} \right)_V
=\frac{1}{c_V V}\left( p-T\left(\frac{\partial p}{\partial T}\right)_V  \right)=-\frac{T}{4V},\eqno{(3)}
$$
where $p$ and $c_{V}$ are pressure and the specific heat at constant volume, respectively.
Then the entropy increase could be interpreted as the free expansion of the radiation due to 
the volume increase by the black hole formation.\cite{rf:SOR}  
It should be noted that the volume within an event horizon for Kerr-Newman solution is almost infinite, 
whereas the surface area is finite.

 This interpretation of the volume increase $V_{\rm BH} \simeq V_{\star} (M/m_{\rm pl})^2$ is 
supplemented by the following consideration.
The uncertainty principle $ \Delta x \Delta p \geq \hbar $ shows one massless particle energy $\Delta e$ 
should satisfy the relation
 $ \Delta e \simeq \Delta p c \geq \hbar c/\Delta x$.  Taking $\Delta x \simeq r_g =2M$, the upper limit of the 
particle number is given by \cite{rf:ZUR, rf:TOM}
  $$
N_{\rm BH} \simeq M/(\Delta e) \leq GM^2/(\hbar c) =(M/m_{\rm pl})^2  .\eqno{(4)}
$$
On the other hand, the number of single particle quantum states in phase space of a black hole is given by 
$$
n_{\rm BH} \simeq V_{\rm BH} (\Delta p)^3/\hbar ^3 \simeq (M/m_{\rm pl})^2  ,\eqno{(5)}
$$
where $V_{\rm BH} \simeq V_{\star} (M/m_{\rm pl})^2$ and $V_{\star} \simeq (\Delta x)^3$ are used.
Then the number of ways $ W$  to distribute 
the $N_{\rm BH}$ particles among the $n_{\rm BH}$ states is
$$
W=(N_{\rm BH} +n_{\rm BH}-1)!/(N_{\rm BH}! (n_{\rm BH}-1)!) .\eqno{(6)}
$$
Taking Stirling's formula $\ln n! \simeq n\ln n-n$, the entropy becomes as 
$$
S =\ln W \simeq (M/m_{\rm pl})^2  \simeq A/l_{\rm pl}^2.\eqno{(7)}
$$

Then, even if an object to become a black hole is a neutron star and the nucleon number $N_{\rm N} 
\simeq M/m_{N} $ is 
much smaller than $N_{\rm BH}$,
the entropy becomes $S \propto  A/l_{\rm pl}^2$, taking into account the increase of the black hole phase space.  
It is the same for the massive star,
where the photon number $N_{\star}$ is of order $N_{\star}\simeq Mc^2/(k_{\rm B}T_{\star}) 
\simeq (M/m_{\rm pl})^{3/2}  $ , 
which is much smaller than $N_{\rm BH}$.

If we consider that a black hole is composed of massless particles with energy $\Delta e$, the temperature is 
on the order of 
$T \simeq 
\Delta e \simeq \hbar c^3/(GM) \simeq m_{\rm pl}c^2(m_{\rm pl}/M) $ \cite{rf: SCA}.  The entropy 
 becomes $ S \simeq M/T\simeq A/l_{\rm pl}^2$ and the effective volume $V_{\rm ef}\propto M^5$ is 
estimated from 
$ S\simeq sV_{\rm ef} \propto T^3V_{\rm ef} \propto V_{\rm ef} /M^3 \propto M^2$,
which also indicates the increase of the effective black hole volume. 
The relation  $V_{\rm ef}\propto M^5$ is also derived from 
$V_{\rm BH}\simeq V_{\rm ef}\simeq (\Delta x)^3 N_{\rm BH} \simeq V_{\star}(M/m_{\rm pl})^2\propto M^5$ .

The problem is the definition and the interpretation of the black 
hole volume.  Taking the appropriate distribution of matter, the 
effective volume of the collapsing object could be enlarged \cite{rf:SOR}.  
For the spherical configuration, the volume in general relativity is estimated as
$$
V_{GR} \simeq 4\pi \intop^{R} {g_{rr}}^{1/2}r^2 dr =4\pi \intop^{R} \frac{r^{1/2}}{\xi ^{1/2}}r^2dr, \eqno{(8)} 
$$
where $g_{rr}=1-2m(r)/r $ and $m(r)$ are the radial metric component in spherical configuration and
the mass within radius $r$,  taking $\xi=r-2m(r)  (>0)$.  The proper distance of the composed matter from its 
Schwarzschild radius is given by 
$$
  D(r)={g_{rr}}^{1/2}\xi=r^{1/2}{\xi}^{1/2} .\eqno{(9)} 
$$
If we take the physical possible condition as 
$$
 D(r) \geq \lambda (r) ,\eqno{(10)} 
$$
where $\lambda$  is the wave length of the matter(radiation), the volume $ V_{GR}$  becomes
$$
V_{GR} \simeq 4\pi \intop^{R} \frac{r^{1/2}}{\xi ^{1/2}}r^2dr \leq 4\pi \intop^{R} \frac{r^3}{\lambda (r)}dr ,\eqno{(11)} 
$$
since $\xi ^{-1/2}\leq r^{1/2}\lambda^{-1}$.  Even if we take $\lambda \simeq l_{pl}$, $V_{GR} \leq R^4 \simeq M^4$. 
 So there is no good ways to derive $V_{GR} \simeq M^5$, before the black hole formation.

  After the black hole formation $( 2m(r) \geq r)$, it is difficult to estimate the volume within the 
horizon, where time and radial coordinates are interchanged.  If we take the time coordinate as spacelike and take 
the evaporation time of the black hole $(t_{ev} \simeq M/(r^2T^4)\simeq M^3\,) \,\,$ \cite{rf: HAW1, rf:ZUR,rf: TOM} as the upper limit of the coordinate $t$, 
the effective volume within the black hole 
is estimated as $V_{BH} \simeq r^2 t_{ev} \simeq M^5$.

 At the fundamental level, black holes are genuine 
quantum objects, where the de-Brogile length corresponding to the black 
hole temperature is comparable to the gravitational length,  it is necessary to include the non local effect 
of the quantum mechanics 
for its entropy interpretation. 

From the above consideration, the microscopic states of black hole entropy could be interpreted 
as the standard statistical state number 
in the phase space and the entropy could be taken as the extensive state quantity, 
even if it is proportional to the horizon surface area, 
if we accept the increase of the effective black hole volume.  The enormous entropy increase and temperature 
decrease
through the black hole formation are
understandable as the free expansion of composed particles to this increased black hole volume.

$ \vspace{5pt} $

\end{document}